\documentclass[aip,jap,preprint]{revtex4-1}  
\usepackage{graphicx}  
\usepackage{dcolumn}   
\usepackage{bm}        
\usepackage{subfigure}
\begin{document}


\title{Radial phononic thermal conductance in thin membranes in the Casimir limit: Design guidelines for devices}

\author{T. A. Puurtinen}
\author{I. J. Maasilta}
 \affiliation{Nanoscience Center, Department of Physics, University of Jyv\"askyl\"a, P. O. Box 35, FIN-40014 Jyv\"askyl\"a, Finland}
 \email{maasilta@jyu.fi }


\begin{abstract}
In a previous publication \cite{casimir1}, we discussed the formalism and some computational results for phononic thermal conduction in the suspended membrane geometry for radial heat flow from a central source, which is a common geometry for some low-temperature detectors, for example. We studied the case where only diffusive surface scattering is present, the so called Casimir limit, which can be experimentally relevant at temperatures below $\sim$ 10 K in typical materials, and even higher for ultrathin samples. Here, we extend our studies to much thinner membranes, obtaining numerical results for geometries which are more typical in experiments. In addition, we interpret the results in terms of a small signal and differential thermal conductance, so that guidelines for designing devices, such as low-temperature bolometric detectors, are more easily obtained. Scaling with membrane dimensions is shown to differ significantly from the bulk scattering, and, in particular, thinning the membrane is shown to lead to a much stronger reduction in thermal conductance than what one would envision from the simplest bulk formulas.     
\end{abstract}
\maketitle

\section{Introduction}

At low temperatures and in nanoscale samples, phonon thermal conduction can be a bit more complex than in macroscopic samples at room temperature, where typically bulk diffusive scattering dominates and Fourier's law is accurate. At low temperatures, the complications arise because at temperatures $T$ much below the Debye temperature $\theta_D$, $T < \theta_D/30$, the typically dominant diffusive phonon-phonon scattering mechanism dies away \cite{bermanbook}. In many materials, then, the remaining bulk scattering mechanisms (various phonon-impurity scattering channels) become so weak that phonon transport becomes ballistic inside the material \cite{zimanbook}. On the other hand, if the dimensions of the device are below typical mean free paths, similar issues arise even at room temperature. It has been observed, for example, that in crystalline silicon, 40 \% of the thermal conductance comes from phonons with mean free path larger than 1 $\mu$m \cite{regner}. This means that for Si membranes with thickness below 1 $\mu$m, a simple bulk diffusive thermal model fails even at room temperature.

When bulk transport is mostly ballistic, one is then left with the question what happens at the surfaces of the sample. Theoretically, this issue was first discussed by Casimir in his seminal work on uni-directional heat flow in rods \cite{casimir}. He analyzed the limit where the rod surfaces where so rough that the surface phonon scattering was fully diffusive, i.e. an incoming phonon would scatter to any direction in half space with equal probability. The result was that one could describe thermal conductance $G$ by the usual scaling law $G=\kappa A/L$, where $\kappa$ is an effective local thermal conductivity, $A$ is the cross sectional area of the rod, $L$ its length. Here, $\kappa$ not only depends on the material parameters, but is also dependent on the geometry of the problem. For example for cylindical rods, one can write $\kappa=\frac{1}{3} C \overline{v} l$, where $C$ is the specific heat capacity of the phonon gas, $\overline{v}$ is a properly defined average speed of sound \cite{zimanbook}, and the effective mean free path $l$ is simply given by the diameter of the rod. Later, the results was generalized for cases, where the surface scattering was only partially diffusive  \cite{bermanziman,zimanbook}, and for more complex cross-sectional shapes and effects of crystalline symmetries \cite{mccurdy, wybourne, eddison, maris}. Note also that the typical Casimir theory does not take into account end effects, in other words effects due to bulk contacts to a rod with a finite length \cite{klitsner}.

In experiments in the beam geometry \cite{panu,fon2,tighe,li,bourg1,bourg2,rostem,osman}, heat flow is indeed uni-directional, and the standard Casimir theory can at least in principle be compared with the experiments. However, in many cases the experimental realization is such that the heat flow is radial (Fig. 1), meaning that there is a central heat source, and heat can spread in all directions. This geometry is particularly apparent in the case of membrane-supported low-temperature bolometric radiation detectors \cite{enss,kimmo}, where the phononic thermal conductance of a suspended membrane is the limiting mechanism for heat dissipation. Experiments on suspended SiN membranes at sub-Kelvin temperatures \cite{zen,jenni,hoevers,holmes} have also clearly demonstrated that the sub-Kelvin thermal conductance in the membrane geometry is typically not limited by bulk scattering and can even approach the fully ballistic limit.  For full understanding of the possibilities of the experiments, one therefore should also study the Casimir limit for the radial heat flow geometry. The theoretical basis for that case was set in Ref. \cite{casimir1}, where also some numerical examples were discussed.

  \begin{figure}
\includegraphics[width=\textwidth]{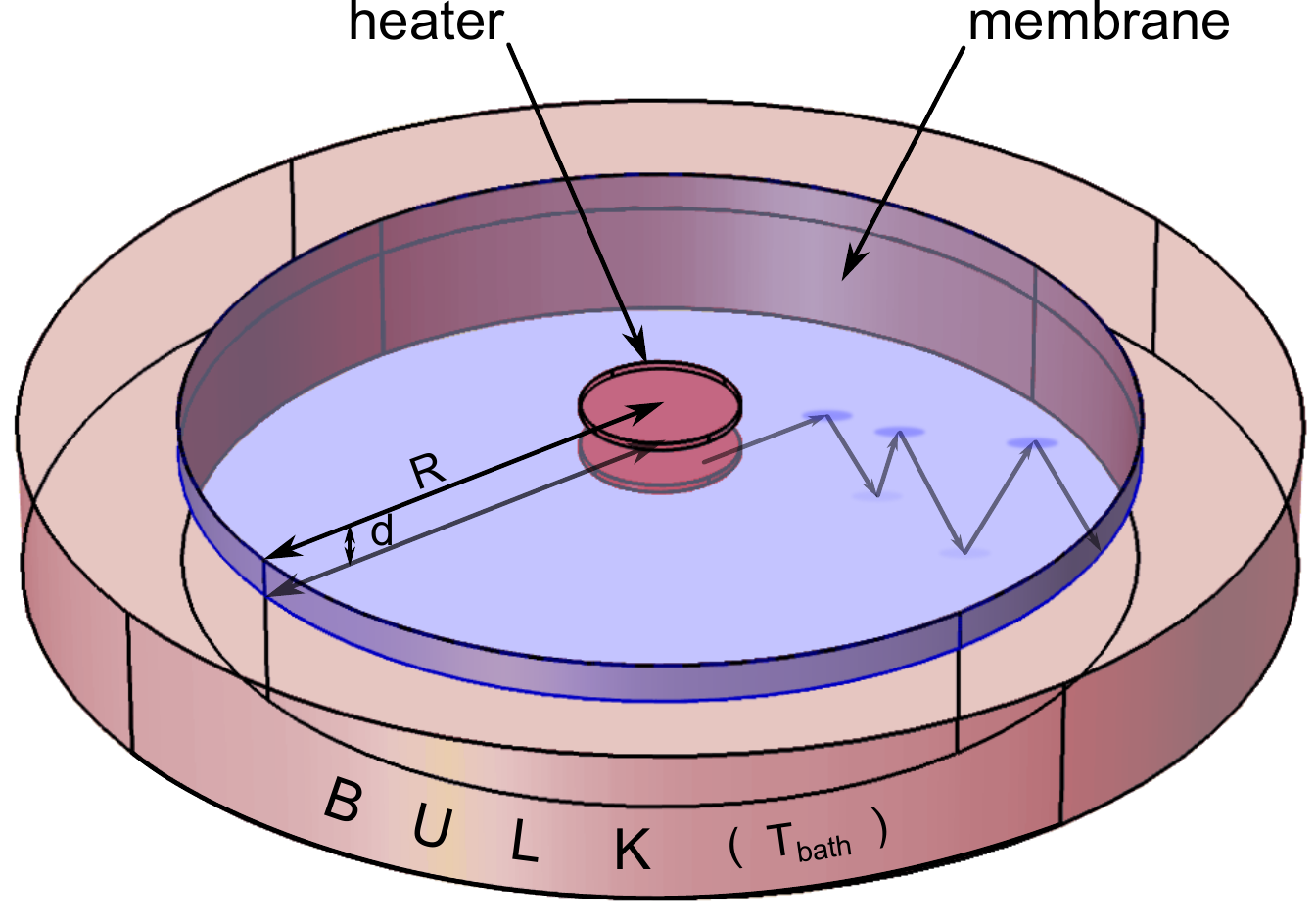}
\caption{[Color online] A schematic drawing of the geometry considered. Heat is generated in a centrally located circular heater, and it is conducted radially outward along a membrane of finite width $d$ with fully diffusive surfaces. The membrane has a finite size defined by a contact to a bulk with constant bath temperature $T_{bath}$.}  
\end{figure}
In this paper, we extend the work initiated in Ref. \cite{casimir1} to geometries with much higher aspect ratio of membrane size to thickness, which means that much more realistic sample geometries are now studied. In addition, we not only calculate temperature profiles, but calculate the thermal conductance, both in small temperature difference and large temperature difference limits. 
The main idea is to give simple phenomenological laws, how the Casimir-limited thermal conductance scales with the geometrical parameters of the problem (membrane thickness, membrane size, heater/detector size), and what is absolute value is. These laws can then be used in practice for detector designs, for example. As before, only numerical solutions are possible, and in contrast to the beam geometry, one cannot use the bulk formulas for thermal conductivity at all in the radial heat flow case. In other words, there is no equivalent general rule that the mean free path would be proportional to some dimension of the sample. 
 
\section{Thermal conductance limited by bulk scattering in the radial flow geometry}

The radial flow geometry is peculiar even in the bulk scattering case   in that the scaling of the linear (small temperature difference) thermal conductance $G$ with sample dimensions is different from the simplest uni-direction flow case, because the cross sectional area is a function of the radial coordinate. From Fourier's law, one can easily derive \cite{vu,leivo,casimir1} that 




\begin{equation}
G=\frac{2\pi d}{\ln(r_{1}/r_{0})}\kappa,
\label{G2d}
\end{equation}     

where $d$ is the membrane thickness, $\kappa$ the bulk thermal conductivity, and $r_{1} > r_{0}$ are the two radii where the two temperatures in $\Delta T$ are defined (for example,  membrane edge radius and heater radius). We see that the scaling with $d$ is still intuitively linear, but with respect to the heater size and membrane size it is logarithmic. It is thus not very easy to decrease $G$ by increasing the membrane size in the radial flow geometry.   



If one assumes some functional form for the temperature dependence of the bulk thermal conductivity, for example $\kappa=\alpha T^m$, one can solve \cite{jenni} the temperature profile $T(r)$ even in the general case with arbitrary input power $P$  where the temperature in the membrane can be much higher than the bath temperature at the membrane edge $T(R)=T_{bath}$ :

\begin{equation}
T(r)=\left [\frac{(m+1)P}{2\pi \alpha d}\ln \left ( \frac{R}{r} \right )+T(R)^{m+1} \right ]^{1/(m+1)} ,
\label{bulkdiff}
\end{equation} 

where $R$ is the radius of the membrane. On the other hand, in the small signal limit where $\Delta T << T_{bath}$, this simplifies to 

\begin{equation}
T(r)= \frac{P}{2\pi d \kappa(T_{bath})}\ln \left ( \frac{R}{r} \right )+T_{bath},
\label{bulkdiff2}
\end{equation}  

where we do not have to consider the explicit form of the temperature dependence of $\kappa$.           
          
\section{Casimir limit in the radial flow geometry: definition of the problem}  

As is detailed in Ref. \cite{casimir1}, the consideration of fully diffusive surface scattering at the surfaces leads to a a fairly simple one dimensional integral equation for the unknown temperature profile $T(r)$, as long as we assume cylindrical symmetry for the heater and the membrane edge, and symmetry between the top and bottom membrane surfaces. For geometries where the size of the membrane is much larger than the size of the heater/detector, the symmetrical solutions are a good approximation to a more realistic case, where only the top surface is heated and where the membrane edge and the heater are not circular. Moreover, it is possible to actually fabricate devices, such as membrane-isolated superconducting X-ray calorimeters, where both the detector element and the membrane edge are circular \cite{xarray}.   


Here, we only consider the case of an isotropic material and thus do not take into account any phonon focusing effects \cite{wolfe,mccurdy} due to crystal symmetries. This is a good approximation especially for amorphous materials such as SiN, which is the most common membrane material. All the material parameters of the problem then combine into one parameter, the phononic Stefan-Boltzmann constant $\sigma$, which is given \cite{klitsner} by 

\begin{equation}
\sigma=\frac{\pi^2k_B^4}{120\hbar^3}\left (\frac{2}{c_t^2}+\frac{1}{c_l^2} \right ),
\end{equation}  

where $c_t$ and $c_l$ are the transverse and longitudinal speeds of sound of the material, respectively. 

The equation to be solved finally reads for the fourth power of the temperature profile $Z(r)=T^4(r)$:

\begin{equation}
Z(r)=\int_0^R\!\!dr_{2}G(r,r_{2})Z(r_{2})+T_{bath}^4H(r)+Cf(r),
\label{eqshort}
\end{equation}

where the kernel $G(r,r_{2})$ is given by

\begin{equation}
G(r,r_{2})= \frac{2d^2r_{2}(r^2+r_{2}^2+d^2)}{\left [ (r^2+r_{2}^2+d^2)^2-4r^2r_{2}^2  \right ]^{3/2}},
\label{kernel}
\end{equation}      

the "edge" function $H(r)$ as

\begin{equation}
H(r)= \frac{1}{2} \left ( \frac{r^2+d^2-R^2}{\sqrt{(r^2+R^2+d^2)^2-4r^2R^2}} +1 \right ),
\end{equation}      

and the constant $C$ is the normalized external power input $C=2q/\sigma$, where $q$ is input power per unit area, so that the function $f(r)$ is one where power is applied and zero where it is not. The free geometrical parameters of the problem, to be varied, are thus a) the membrane thickness $d$, b) the membrane size (radius) $R$, and c) the size of the heater defined by the function $f(r)$. 

\section{Casimir limit in the radial flow geometry: numerical results}

\subsection{General considerations}

Equation \ref{eqshort} has to solved numerically, as no obvious analytic solution exists. Using the terminology of mathematical literature, equation \ref{eqshort} can be classified as a linear Fredholm equation of the second kind \cite{numrec}. 
We solve it by using the Nystr\"om method \cite{numrec}, which uses the Gauss-Legendre quadrature rule for discretization of the integral, and triangular decomposition techniques for the inversion of the obtained linear equations. By trial and error we have seen that the details of the problem influence how many points are needed in the discretization. The higher the aspect ratio between membrane radius and thickness, the higher number of points must be used. Here, we went up to 10 000 points  (10 000 by 10 000 matrix inversion) to get accurate results for the thinnest membranes studied here (aspect ratio $R/d=2000$).  

In all the results that follow, we use real units and realistic parameter values. However, as was noted before \cite{casimir1}, the problem is scalable in the sense that if all the dimensions are scaled, the result for the temperature profile stays the same for a given input power. The material parameters used are for SiN, with $c_t=5726$ m/s and $c_l=9669$ m/s calculated from literature values  \cite{young} for the Young's modulus $Y=250$ GPa, Poisson ration $\nu=0.23$ and density $\rho= 3100$ kg/m$^3$ .  

\begin{figure}
\includegraphics[width=\textwidth]{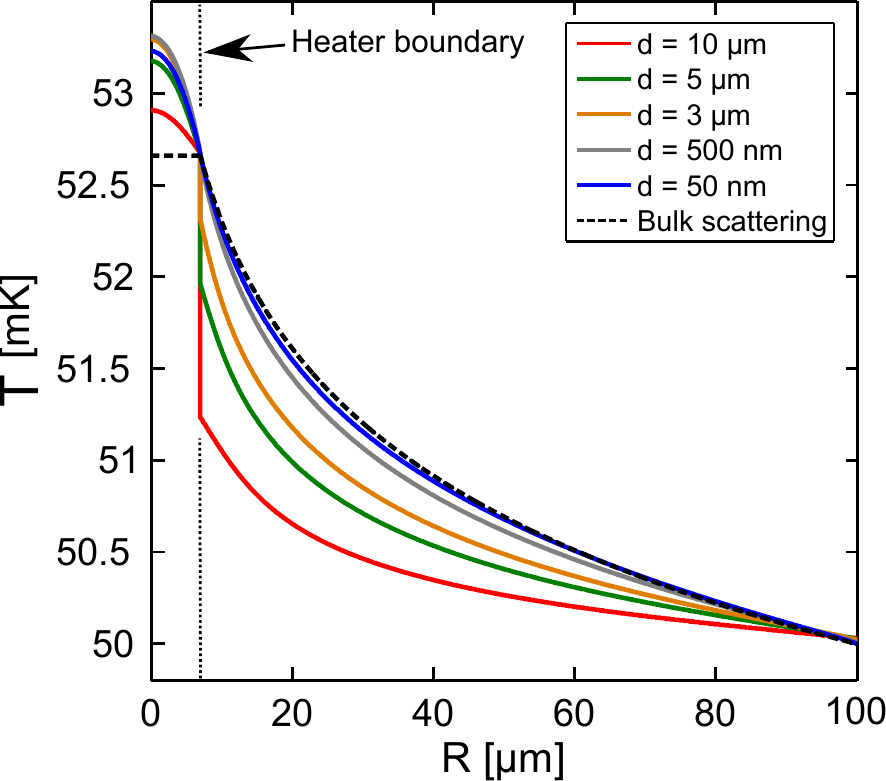}
\caption{[Color online] Calculated temperature profiles for radial Casimir heat conduction in SiN (lines) for $R=100 \mu$m and $d$ varying from 50 nm to 10 $\mu$m, with radius of heater 7 $\mu$m and input power scaled for each curve in such a way that the temperature at the heater edge is constant. Higher curves have lower $d$, and bath temperature was set to 50 mK. Dashed line shows the bulk diffusive result, Eq. \ref{bulkdiff2}.} \label{fig1}
\end{figure}

As an example, we plot some representative temperature profiles in Fig. \ref{fig1} for different membrane thicknesses, where the heater radius equals 7 $\mu$m and the membrane edge is at $R= 100 \mu$m, and $T_{bath}$ was set to 50 mK. Here, we used only small input heating power such that we are in the linearized regime where $\Delta T << T_{bath}$. By comparing the results to the simple bulk diffusive limit, Eq. \ref{bulkdiff2}, we see that the thinner the membrane gets, the closer the temperature profile resembles the bulk scattering limit. However, there is still an observable deviation between the Casimir and bulk cases, even for the thinnest membranes $d=50$ nm calculated. Even larger scale calculations would be required for this membrane size to see if the bulk limit is really reached for some finite $d$. Nevertheless, we can see that the deviations in the 50 nm thick membrane case are quite small so that the temperature profile {\em outside} the heater could be approximated by the simple form, Eq. \ref{bulkdiff2}. However, the temperature profile inside the heater is not the same (bulk model assumes a constant), and this will still affect the computed values for thermal conductance $G$, as will be seen below. 



\subsection{Thermal conductance}

As long as the temperature profiles do not agree with the bulk limit, we cannot easily define a useful effective thermal conductivity $\kappa$, as is the case with Casimir limit in the rod geometry. (If the temperature profile shapes in the Casimir limit and in the bulk case were equal, we could simply do a fit for each $d$, and extract a $d$-dependent effective $\kappa$.) Thus we are limited to calculating the thermal conductance $G$, and then studying how it scales with the dimensions of the problem. 

We thus define a linear thermal conductance $G$ by the relation $P=G\Delta T$, where $P$ is the total input power in the heater, and $\Delta T$ is the temperature difference between the heater temperature and the bath temperature. For this definition, we require that $\Delta T << T_{bath}$, so it is only valid for small input powers. Moreover, we have to specify what we mean by the heater temperature. As we can see form Fig. \ref{fig1}, the temperature within the heater region is not necessarily constant. We therefore define the heater temperature to be the average over the heater region.

Another choice for the definition of thermal conductance is the differential one: $G_{d}=dP/dT$, where the derivative is taken with respect to the heater temperature. This is a well defined object for all heater temperatures, and therefore the only one that is relevant in the cases where $\Delta T$ is of the same order of magnitude or larger than $T_{bath}$. This is a typical situation for bolometric low-temperature radiation detectors, such as transition-edge sensors, which utilize the so called negative electrothermal feedback to improve detector performance \cite{enss}.       

\begin{figure}
\includegraphics[width=\textwidth]{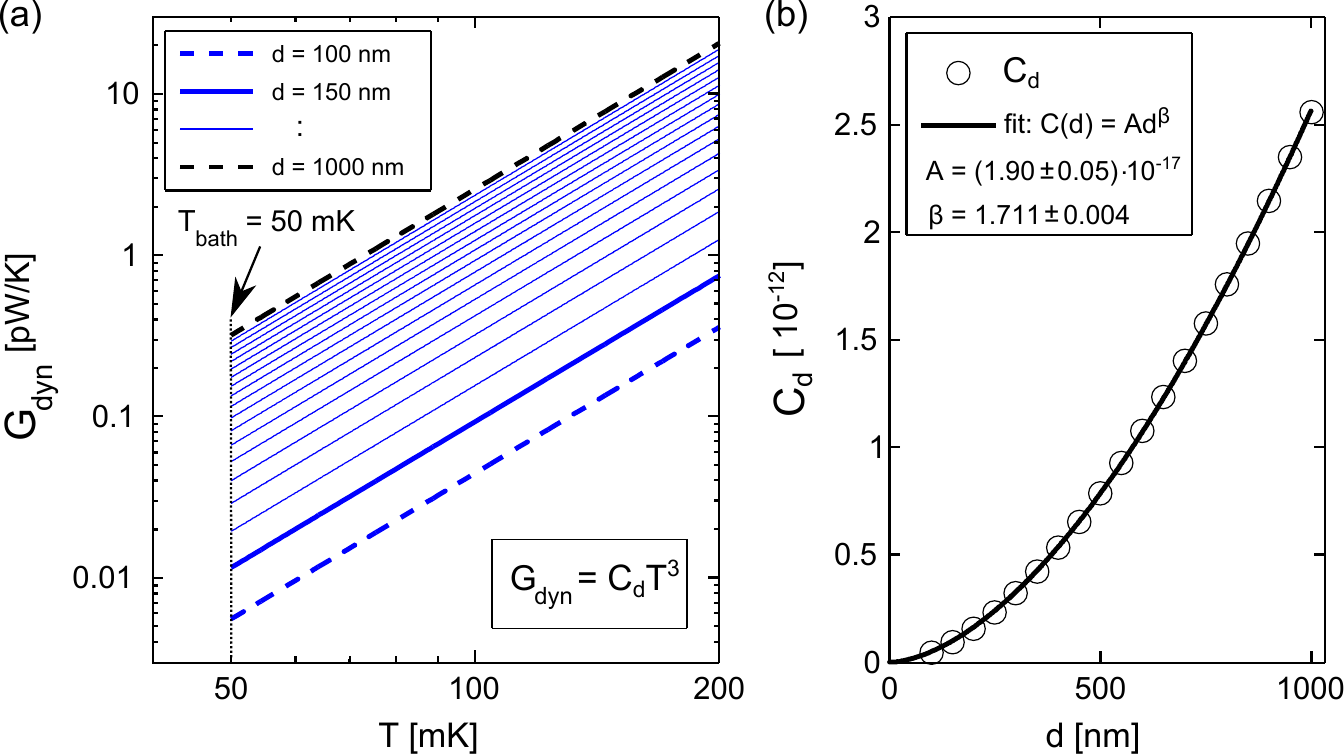}
\caption{[Color online] (a) Calculated differential conductances $G_d=dP/dT$ vs heater temperature $T$ for radial flow Casimir heat conduction (lines) for $R=100 \mu$m and varying $d$, in log-log scale. All curves follow a power law $G_d=C_dT^3$. (b) The dependence of of the prefactor $C_d$ for $T= 100$ mK.  Radius of heater was 7 µm, and bath temperature 50 mK. } \label{fig2}
\end{figure}

In Fig. \ref{fig2} we show examples of the calculated differential thermal conductances $G_d$ as a function of heater temperature, for varying membrane thicknesses. 
For all values of $d$, the temperature dependence follows the power law $G_d=C_d T^{3}$, which is a direct consequence of the phonon Stefan-Bolzmann law $\sigma T^4$ for emitted radiative power/unit area  from a surface element \cite{casimir1}. However, more interesting is how the prefactor $C$ varies as a function of the geometry. An example of that is shown in Fig. \ref{fig2} (b), where the prefactor $C$ is plotted as a function of $d$ for heater temperature $T=$ 100 mK. It is clearly non-linear with $d$.

To simplify the discussion, all the results in the following will be shown only for the linear $G$, which does not really limit the generality of the results. That can be understood from Fig. \ref{fig2} (a): the linear $G$ is just the $T \rightarrow T_{bath}$ limit of all the $G_{d}$ vs. $T$ curves, and as the temperature dependence is exactly the same for all geometries, the linear $G$ results can always be scaled up to give $G_d$ results by multiplying with the factor $(T/T_{bath})^3$.

\begin{figure}
\includegraphics[width=\textwidth]{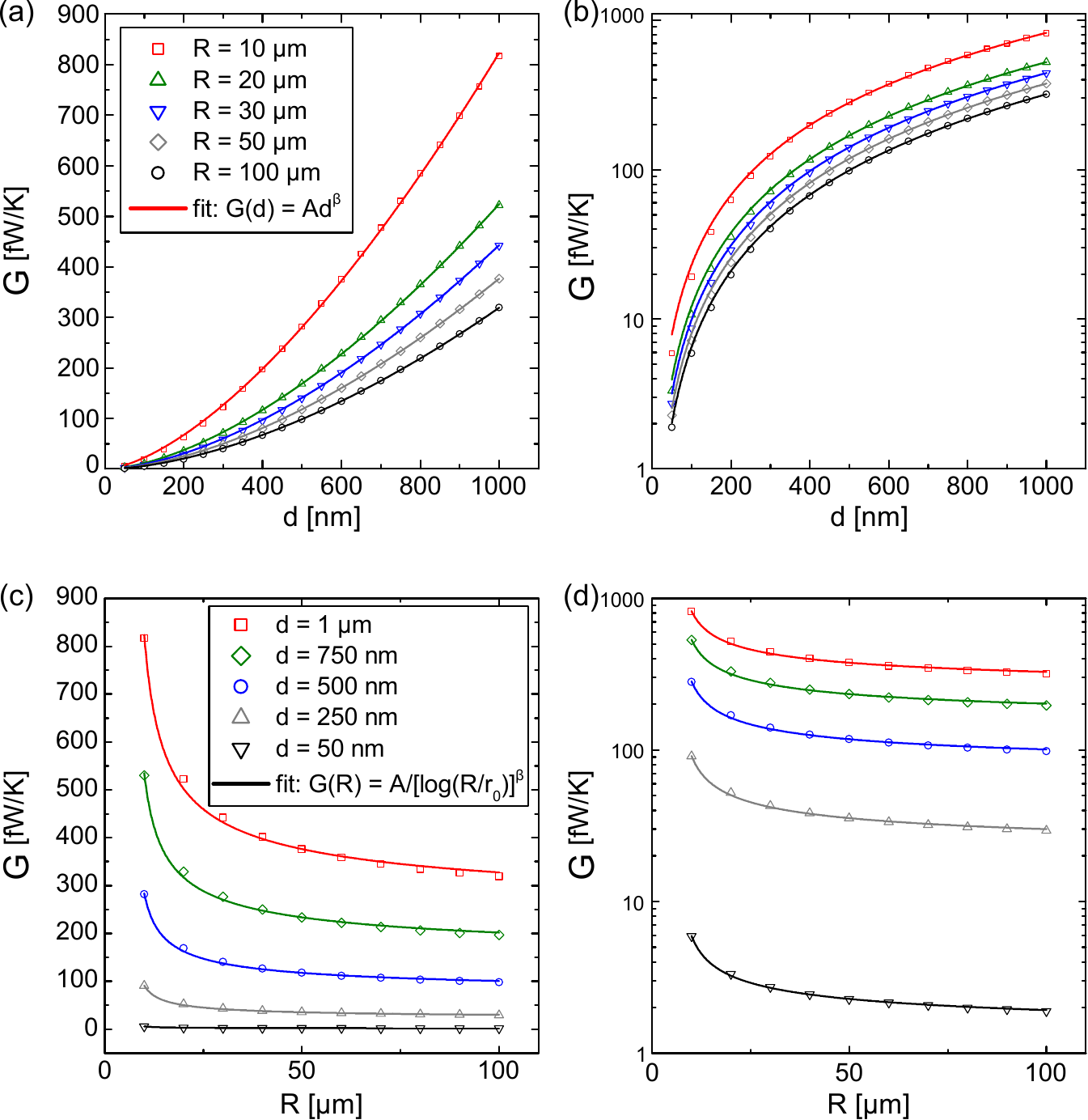}
\caption{[Color online] Calculated linear thermal conductances $G$ for radial flow Casimir heat conduction (symbols),  as a function of membrane thickness $d$ in (a) linear and (b) log-linear scales, with varying membrane sizes $R$.  (c) $G$ as a function of membrane size $R$in (c) linear and (d) log-linear scales with varying membrane thickness $d$. $T_{bath}$ = 50 mK. The lines are two-parameter fits to functions of the form  $Ad^{\beta}$ in (a) and (b), and $A/(\ln(R/r_0))^{\beta}$ in (c) and (d), where $r_0$ is the heater radius.} \label{fig3}
\end{figure}

The interesting issue is now: how does the linear thermal conductance $G$ scale with $d$ and $R$? Fig. \ref{fig3} presents those results for the case $T_{bath}=$ 50 mK. We see, first of all, that {\em $G$ is always non-linear function of the membrane thickness $d$}, in contrast to the bulk result Eq. \ref{G2d}, which shows a linear dependence on $d$. We have fitted simple functions of the form $Ad^\beta$ to the calculated data in Fig. \ref{fig3} (a), with a surprisingly good quality of the fits. The results of the fits are shown in Table \ref{table1}. The exponent $\beta$ ranges between 1.5-1.7 for these cases, showing that $G$ increases with $d$ more strongly than for the bulk scattering case. In other words, to increase or decrease $G$, one needs a smaller change in $d$ what one would expect from the simple minded extrapolation of the bulk scattering case. 
It is somewhat surprising, that scattering from surfaces be more effective than scattering from bulk.       

\begin{table}[h]
\begin{center}
\begin{tabular}{rcr}
$R$ [$\mu$m] & $A$ [mW/(Km$^{\beta})]$ & $\beta$ \\
\hline
100                                                   & $4.39$  & 1.690 \\
90                                                     & $4.33$  & 1.687 \\
80                                                     & $4.26$  & 1.684 \\
70                                                     & $4.18$  & 1.680 \\
60                                                     & $4.07$  & 1.676 \\
50                                                     & $3.94$  & 1.670 \\
40                                                     & $3.75$  & 1.662 \\
30                                                     & $3.46$  & 1.649 \\
20                          & $3.25$  & 1.632 \\
10                          &                        $1.67$  & 1.551 \\
\hline
\end{tabular}
\end{center}
\caption[Fitting parameters for $G(d)$]{Fitting parameters $A$ and $\beta$ for $G(d)=Ad^\beta$ with variable radius $R$.}
\label{table2}
\end{table}

\begin{table}[h]
\begin{center}
\begin{tabular}{rlr}
$d$ [nm] & $A$ [pW/K] & $\beta$ \\
\hline
1000                    &                          $\quad 3.499$  & 0.4591 \\
900                      & $\quad 2.938$  & 0.4689 \\
800                                                   & $\quad 2.413$  & 0.4793 \\
700                                                   & $\quad 1.927$  & 0.4904 \\
600                                                   & $\quad 1.482$  & 0.5023 \\
500                                                   & $\quad 1.085$  & 0.5153 \\
400                                                   & $\quad 0.7383$ & 0.5296 \\
300                                                   & $\quad 0.4475$ & 0.5456 \\
200                                                   & $\quad 0.2200$ & 0.5633 \\
100                                                   & $\quad 0.06572$& 0.5780 \\
50                                                     & $\quad 0.02086$& 0.5606 \\
\hline
\end{tabular}
\end{center}
\caption[Fitting parameters for $G(R)$]{Fitting parameters $A$ and $\beta$ for $G(R)=A/(\ln(R/r_0))^{\beta}$ with variable thickness $d$.}
\label{table1}
\end{table}

As a function of the membrane radius $R$ [Fig. \ref{fig3}(b)], the situation is a bit different in the sense that the dependence is quite weak. That is of course to be expected due to the radial nature of the problem, as even in the bulk case the dependence is logarithmically slow. Nevertheless, the functional dependence is different from the bulk case, best fitted by functions of the type $A/(\ln(R/r_0))^{\beta}$, where $r_0$ is the heater radius and exponent $\beta$ ranges between 0.46-0.56. This means that the dependence is again stronger than in the bulk case, but still not very effective if one increases the membrane size to much more than twice the heater size. The results of the fits are again listed in Table \ref{table2}.



\section{Discussion and conclusions}

We studied computationally phonon thermal conduction in the diffuse surface scattering limit (Casimir limit) for the geometry where heat flows radially out of a central heater into a suspended membrane of finite size. In contrast to the Casmir limit for the unidirectional heat flow (beam geometry), no simple results are available in terms of an effective size dependent thermal conductivity, as no correspondence can be drawn between the bulk scattering and surface scattering results. Thinner Casmir-limit membranes do seem to slowly approach the bulk case, which can be intuitively understood from the point of view that the effect of the ballistic inner part of the membrane gets smaller and smaller as the membrane gets thinner. 

Here, the main focus of the work was to calculate cases that are realistically thin, and analyze the results in terms of how the thermal conductance $G$ (extensive property) scales with the sample dimensions. This way, guidelines for real thermal conductance-related problems, such as the proper design of bolometric low-temperature detectors, can be presented. Here, we specifically set the bath temperature (50 mK) and dimensions to correspond to realistic detector designs, and gave all results in real units, although the problem itself is actually scalable with respect to the dimensions. 

Typically, the detector/heater size is fixed by other considerations. Thus we simply concentrated in studying the dependence of $G$ on the membrane thickness and the membrane size. In both cases, the results showed that the dependence is nonlinear in both cases, and stronger than in the bulk scattering limited radial flow problem. This has clear implications for possible detector applications: i) Membrane size is useful for the control of $G$ only in the size range where it is approximately twice the size (in linear dimension) of the heater. Beyond that, the achieved decrease in $G$ is small. ii) Membrane thickness is a much more effective parameter than in the bulk case. In fact, the results showed that it was possible to reach quite low values of $G \sim 10$ fW/K with still reasonably thick membranes of thickness 50 nm. This level of $G$ is usually achieved with the help of long and narrow beams \cite{kenyon,goldie} or with nanoscale hot-electron systems \cite{wei,govenius}. We therefore propose that a roughened but thin full membrane can possibly do the same job, but it would be mechanically more robust (no narrow beams) and would not require nanolithography to create a small electron volume.

\section*{Acknowledgements}
 This research was supported by Academy of Finland project number 260880.

\end{document}